\documentclass{ws-procs975x65}

\begin{document}

\title{ Search for the $\Theta^+$ in Photoproduction on the Deuteron 
\footnote{\uppercase{T}he g10 experiment at 
\uppercase{J}efferson \uppercase{L}ab: 
\uppercase{H}icks and \uppercase{S}tepanyan, co-spokesmen.}
}

\author{K.~H. Hicks\footnote{\uppercase{T}his work is supported in part by the 
\uppercase{N}ational \uppercase{S}cience \uppercase{F}oundation}
~for the CLAS Collaboration}

\address{Department of Physics and Astronomy \\
Athens, OH 45701, USA \\ 
E-mail: hicks@ohio.edu}

\maketitle

\newcommand{\thp}{$\Theta^+$ }
\abstracts{
A high-statistics experiment on a deuterium target was performed 
using a real photon beam with energies up to 3.6 GeV at the CLAS 
detector of Jefferson Lab.  The reaction reported here is for 
$\gamma d \to p K^- K^+ n$ where the neutron was identified using 
the missing mass technique.  No statistically significant narrow 
peak in the mass region from 1.5-1.6 GeV was found.  An upper 
limit on the elementary process $\gamma n \to K^- \Theta^+$ was 
estimated to be about 4-5 nb, using a model-dependent correction 
for rescattering determined from $\Lambda$(1520) production. 
Other reactions with less model-dependence are being pursued.
}

\section{ Introduction and Results }

The search for pentaquarks, made from four quarks and one antiquark, 
has captured the interest of the nuclear-particle physics community 
since the announcement of a possible experimental signal by the 
LEPS Collaboration\cite{leps}.  Since then, there have been many 
results published, some positive and some null, and the reader is 
referred to a recent review\cite{hicks} for more details.

Here, we focus on the reaction $\gamma d \to p K^- K^+ n$ which was 
previously published\cite{stepanyan} but with low statistics.  The 
present results are for a high-statistics experiment, known as ``g10", 
carried out using the same detector, the CEBAF Large Acceptance 
Spectrometer (CLAS) at Jefferson Lab.  The experimental setup is the 
same as Ref. \refcite{stepanyan}, known as ``g2a",  except for two items: 
(1) the beam energy 
was increased, allowing photons from 0.9-3.6 GeV; (2) the target 
was moved upstream by 25 cm to increase the acceptance for negative 
particles.  The data analysis and event selection cuts used in the 
present analysis are the same as Ref. \refcite{stepanyan}, and the 
photon energy range has been restricted (by software) to match as 
closely as possible the conditions of Ref. \refcite{stepanyan}. In this 
sense, the analysis is a ``blind" analysis, so that no bias was 
introduced in the high-statistics result.

In the g10 experiment, the data was taken at two magnetic field 
settings of the CLAS torus coils.  At both field settings, a luminousity 
of about 25 pb$^{-1}$ was collected, which is nearly 10 times the 
luminousity of the previous g2a data\cite{stepanyan}.  Only the high-field 
data will be presented, which matches the conditions of the g2a data,
although the results from the low-field setting are found to be similar. 
After restricting the photon energy range to be the same as g2a, 
the g10 experiment had 5.9 times the luminousity of the g2a experiment. 
A comparison of the two experiments is shown in Fig. 1, where the 
missing mass of the $pK^-$ system (equal to the mass of the $nK^+$ 
system, from possible \thp decay) is plotted.  The vertical scale 
shows the number of counts in the published g2a data\cite{stepanyan} 
and the g10 data have been scaled by the luminosity, shown by the solid 
histogram.

\begin{figure}[hp]
\centerline{\epsfxsize=10cm\epsfbox{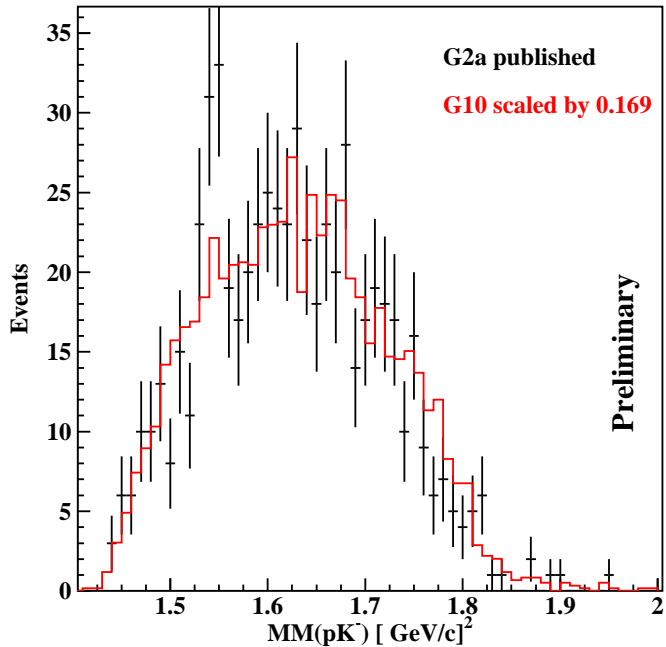}}   
\caption{ Missing mass of the $pK^-$ system for the reaction 
$\gamma d \to p K^- K^+ n$ measured at CLAS.  The points with error 
bars is from Ref. [3], and the solid histogram shows 
the present (high-statistics) results, scaled down by the factor shown.
}
\label{fig:compare}
\end{figure}

Clearly, the peak seen in the g2a data is not reproduced by the g10 data. 
Using the g10 data as a guide to the background shape, the probability 
of a fluctuation of the amount seen at 1.54 GeV in the g2a data is
found to be about 3-$\sigma$ (three standard deviations). The claim 
of a 5-$\sigma$ statistical significance in Ref. \refcite{stepanyan} was 
due to a lower estimate of the background.  We note that the g2a data 
fluctuate downward from the g10 shape on either side of the 1.54 GeV 
``peak".  In hindsight, we see that the evidence for the \thp claimed 
in Ref. \refcite{stepanyan} is due to a combination of an underestimate 
of the background shape and a statistical fluctuation in the region of 
1.54 GeV.  These results show the importance of high statistics, 
along with a ``blind" analysis procedure where the event selection 
criteria are determined before the experiment is done.  

It is now a straight-forward procedure to fit the g10 mass spectra 
with an overall background shape (using a third order polynomial). 
Using a fixed background and fitting a gaussian (with a 6 MeV width, 
equal to the CLAS resolution) across the mass spectrum, an upper limit on 
the number of counts in the mass region of 1.54 GeV is found.  Using the 
luminosity, along with and the gaussian fit results and a detector 
acceptance from Monte Carlo, an upper limit on the {\it measured} 
reaction on deuterium has been calculated.  We assume a uniform 
angular distribution for $\Theta^+$ production, even though the 
CLAS detector does not measure particles at forward angles (the angle 
is momentum-dependent but roughly 15$^\circ$-20$^\circ$ lab for 
$K^-$ and roughly 8$^\circ$-10$^\circ$ lab for the $K^+$). This may 
not be a valid assumption if the $\Theta^+$ is produced primarily 
at forward angles, as suggested by the LEPS data\cite{leps}.

An upper limit on the cross section for the elementary 
reaction $\gamma n \to K^- \Theta^+$ is desired. Of course, the reaction 
we measured was on deuterium, not a free neutron. In order to convert 
from the measured result to the elementary reaction, a theoretical 
model must be used.  The model is complicated by the fact that the 
proton is detected in CLAS, which requires it to have a momentum of 
$>350$ MeV to exit the liquid deuterium target.  
Ideally, the proton in the deuterium target would be a 
spectator to the elementary reaction, having nearly zero momentum 
(smeared by Fermi momentum).  In the g10 experiment, the proton must 
gain momentum by final state (rescattering) reactions.  In order to 
estimate the rescattering correction, we look to the mirror reaction 
$\gamma p \to K^+ \Lambda(1520)$.  In this mirror reaction, the 
neutron would be a spectator, and its momentum is found in g10 by the 
missing momentum.  By cutting on the neutron momentum above  350 MeV, 
the rescattering probability of the mirror reaction is found to be 
about $0.10 \pm 0.01$. Assuming a similar correction for rescattering 
in \thp production on deuterium, the cross section for the elementary 
process is estimated at 4-5 nb.  

The model dependence in the above cross section estimate is 
undesirable but unavoidable.  One could imagine other ways to do the 
rescattering correction, such as using the tail of the Fermi momentum 
above 350 MeV for the proton in deuterium.  In this case, the upper 
limit is increased by a factor of 5, to 20-25 nb.  
Both estimates, one using the $\Lambda(1520)$ model and the other 
using the Fermi tail, are shown as a function of the $nK^+$ mass 
in Fig. 2.  Other models might 
suggest a bigger rescattering probability, thus reducing the upper 
limit.  Clearly, a measurement without a rescattering correction would 
be better.  For example, the reaction $\gamma d \to K^- \Theta^+ p$ 
where the proton is not detected, and the decay $\Theta^+ \to K^0 p$  
is measured, has less model dependence to deduce the elementary 
reaction cross section.  Further analysis of the g10 data is in 
progress and more results are expected soon.

\begin{figure}
\centerline{\epsfxsize=10cm\epsfbox{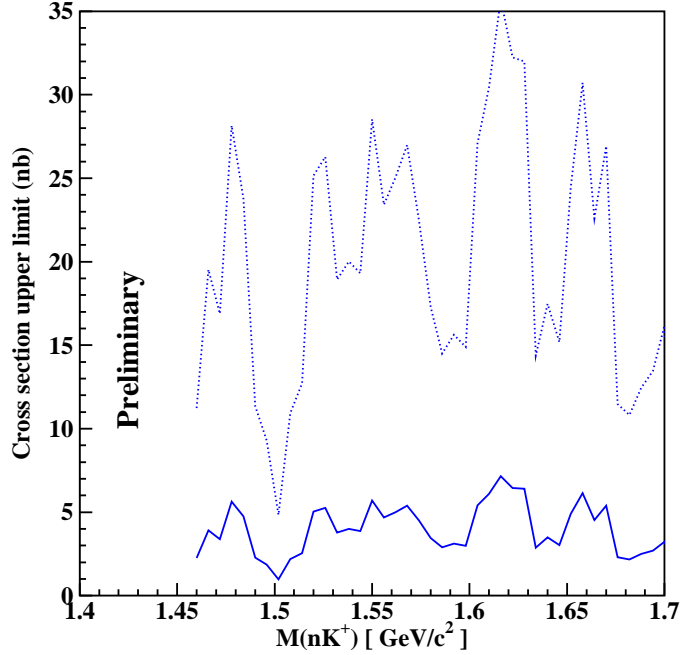}}   
\caption{ Preliminary upper limit for the elementary reaction 
$\gamma n \to K^-\Theta^+$ 
using the $\Lambda(1520)$ for the rescattering correction (lower line) and 
the Fermi momentum tail for the correction (dotted upper line).  All curves 
come from fitting a fixed-width gaussian on top of a fixed polynomial 
background, using the same mass spectrum from the g10 experiment. 
}
\label{fig:upper}
\end{figure}

\section{Summary}

The exclusive reaction $\gamma d \to p K^- K^+ n$ was measured at CLAS
with high statistics. 
No evidence for a narrow peak in the $nK^+$ mass spectrum was 
observed, contrary to earlier low-statistics results\cite{stepanyan}. 
A {\it model dependent} upper limit for the cross section
in the elementary reaction $\gamma n \to K^- \Theta^+$ was estimated 
to be about 4-5 nb, using the $\Lambda$(1520) as a model for the 
rescattering.

\newcommand{\etal}{ {\it et al.}, }

\end{document}